\documentclass[11pt,titlepage]{article}
\usepackage{setspace}
\usepackage{epsfig}
\usepackage{ae}
\usepackage{aecompl}
\usepackage{amsmath}
\usepackage{subfig,graphicx}
\usepackage[round,numbers,sort&compress]{natbib}
\begin{document}
\title{Model plasma membrane indicates the inner leaflet is poised to 
initiate compositional heterogeneities}
\author{
 H.  Giang and   M. Schick}

\date{\today}
\maketitle
\begin{abstract} 
We investigate a model of an asymmetric bilayer consisting of 
sphigomyelin, phosphatidylcholine, and cholesterol in the outer leaflet, 
and phosphatidyl\-ethanolamine, (PE), phosphatidylserine, and cholesterol 
in the inner leaflet. Composition fluctuations are coupled to membrane 
fluctuations via the bending energy which depends upon the local 
spontaneous curvature dominated by the PE in the inner leaflet. This 
brings about a microemulsion in that leaflet with a characteristic 
wavelength of 37 nm as shown by the PE-PE structure function. Thus the 
inner leaflet will respond to an external perturbation most strongly at 
this length.  However the correlation length is approximately 1 nm so that 
composition variations are not seen in the inner leaflet itself. Depending 
upon the strength of the coupling between leaflets, the outer leaflet is 
either a normal fluid or is also a microemulsion. In this model then, the 
lipids of the plasma membrane would not evince inhomogeneities, but would 
be primed to create them from the inner leaflet in response to non-lipid 
perturbations.
\end{abstract}

\section{Introduction}
The hypothesis that the distribution of lipids of the plasma membrane is 
not homogeneous, but rather is characterized by inhomogeneities, denoted 
``rafts'', that can serve as platforms for protein function continues to 
draw much attention.  However much of the evidence for such 
inhomogeneities is indirect \cite{lingwood10}. Furthermore there is no 
agreement on a physical mechanism that would bring about such 
inhomogeneities, although several have been proposed.  If proteins attract 
some lipids in preference to others, then the aggregation of proteins 
would be reflected in an inhomogeneous distribution of lipids. Other 
mechanisms which do not require the presence of proteins have also been 
proposed.  For example, an excess of lipids in one region could result 
from liquid-liquid phase separation \cite{schroeder94}. Indeed {\em in 
vitro} ternary mixtures of a saturated lipid, representing sphingomyelin 
(SM) in the plasma membrane, an unsaturated lipid, like 
phosphatidylcholine (PC), and cholesterol do undergo such a phase 
transition at biologically relevant temperatures \cite{veatch05}. Large 
macroscopically phase-separated regions have, in fact, been observed in 
yeast vacuoles \cite{toulmay13}, but never in plasma membranes.  This may 
be due to the fact that the inner leaflet of the plasma membrane consists 
almost entirely of unsaturated lipid so that there is little tendency for 
phase separation of unsaturated and saturated lipids \cite{wang00}.
 It could then be argued that in the plasma membrane the critical 
temperature for phase separation is below biological temperatures, and 
that the inhomogeneities are due to fluctuations of a nearby critical 
point \cite{matcha11}. If so, one has to understand why so many different 
systems are just poised at the necessary distance from a critical point 
that the fluctuations are just of the size thought to characterize rafts, 
between 10 and 100 nm.  Further, as noted above, there is little tendency 
for the inner leaflet to undergo phase separation at all. A third line of 
thought is that the inhomogeneities indicate that the lipid membrane is a 
two-dimensional microemulsion, a fluid with dynamic but well-defined 
structure. Three-dimensional microemulsions, often consisting of oil, 
water, and an amphiphile are, of course, very well-known \cite{gompper94}.  
The question then arises as to what molecule plays the role of an 
amphiphile in the plasma membrane. Safran and co-workers 
\cite{brewster09,palmieri13} have argued that the unsaturated lipids can 
play such a role by orienting themselves at, or near, the interface 
between saturated lipid-rich and saturated lipid-poor regions. There is 
experimental evidence, however, that this mechanism is not the origin of 
compositional inhomogeneities \cite{heberle13}. That microemulsions could, 
nonetheless, be the source of compositional variations was noted by one of 
us \cite{schick12} observing that microemulsions can be produced by 
mechanisms that do not require an amphiphile. Any mechanism that can 
produce modulated phases can produce a microemulsion as the latter can be 
thought of as resulting from the melting, or disordering, of the former.  
Modulated phases have indeed been observed in mixtures of three, or four 
lipids \cite{konyakhina11,goh13,stanich13} and in yeast vacuoles 
\cite{toulmay13} which renders plausible the idea that the plasma membrane 
might be characterized as a microemulsion.

Two mechanisms that could bring about modulated phases and microemulsions 
without an amphiphile have been proposed. Recently Amazon et al. 
\cite{amazon13} noted that a membrane bending modulus which varied with 
the local composition could bring about modulated phases on a closed 
bilayer vesicle.  Earlier, Schick \cite{schick12} noted that there was a 
significant difference between the large spontaneous curvature of PE and 
the relatively small one of PS, and therefore proposed that the coupling 
of membrane fluctuations and a compositionally-dependent spontaneous 
curvature \cite{leibler87} could be the mechanism at work.  Just such a 
cholesterol-dependent spontaneous curvature of PE was recently invoked by 
us \cite{giang14} to explain the several observations that cholesterol is 
located preferentially in the cytoplasmic leaflet of the plasma membrane 
\cite{brasaemle88,schroeder91,wood90,igbavboa96}. It is this mechanism 
that we shall invoke in this paper.

Because this significant difference in spontaneous curvature of the 
components is only characteristic of the cytoplasmic layer, we must treat 
an asymmetric bilayer, as had been done earlier by Shlomovitz and Schick 
\cite{shlomovitz13}. In that paper, however, the role of cholesterol was 
ignored, and the system was characterized by one order parameter in each 
leaflet. In this paper, we include cholesterol explicitly and characterize 
each leaflet by three components; SM, PC, and cholesterol in the outer 
leaflet, PE, PS, and cholesterol in the inner leaflet. Furthermore we 
require that the chemical potentials of the cholesterol in each leaflet be 
equal reflecting the rapid exchange of cholesterol between leaflets 
\cite{lange81,muller02}. Our model is reviewed in the next section.  We 
calculate structure functions between various components, in particular 
the PE-PE and SM-SM structure functions. We find that over a wide range of 
compositions of the inner leaflet, the PE-PE structure functions have a 
maximum at a non-zero wave number, a feature characteristic of a 
microemulsion. The wave number corresponds to a length of about 37 nm. The 
behavior of the structure function indicates that the inner leaflet 
evinces the largest response at this distance from a perturbation. The 
SM-SM structure function shows that the outer leaflet is either a normal 
fluid or a microemulsion depending upon the strength of the coupling 
between the two leaflets. We also examine the two-particle PE-PE 
correlation function of the inner leaflet obtained from the structure 
function. It has the form of an exponentially damped oscillatory function. 
However the correlation length, which sets the scale of the exponential 
damping, is in general considerably smaller than the wavelength of the 
oscillations. As a consequence, the only observable remnant of the 
oscillations is that the correlation function decays to zero from negative 
values. Consequently, while the system is quite susceptible to 
perturbations of a large size, it does not, of itself, display 
inhomogeneities at such a length scale.

\label{sec:intro}

\section{Materials and Methods}
We consider a bilayer which contains PS, PE, and cholesterol in the inner 
leaflet, and SM, PC, and cholesterol in the outer leaflet. Further, as we 
know that cholesterol tends to order the chains of the other lipids, 
particularly the saturated chains of SM, we consider there to be two 
populations of SM; those with more-ordered chains, SM$_{or}$, and those 
with less-ordered chains, SM$_{dis}$. These populations freely convert 
between one another. This description can produce 
\cite{putzel11,almeida11} the closed-loop phase diagram observed in some 
ternary mixtures \cite{veatch06}. We denote the number of molecules of 
each component in the inner leaflet by $N_{PE}$, $N_{PS}$ and $N_{C_i}$, 
and the total number of molecules in the inner leaflet by $M_I$. Similarly 
denote the number of molecules of each component in the outer leaflet by 
$N_{SM_{or}}$, $N_{SM_{dis}}$, $N_{PC}$, and $N_{C_o}$ and the total 
number of molecules in the outer leaflet by $M_O$. Then the mol fractions 
of the $i$'th component is $x_i=N_i/(M_I+M_O).$ It is more convenient to 
work with the mol fractions defined in each leaflet separately rather than 
in the total bilayer. Thus we introduce the mol fractions of the $j$'th 
component in the inner and outer leaflets,
\begin{align}
	y_{j} =
	\begin{cases} 
		\frac{N_j}{M_i}, & \mbox{if } j=PS,PE,C_i,\\
		\frac{N_j}{M_o}, & \mbox{if } j=SM_{or},SM_{dis},PC,C_o.	
	\end{cases}  
\end{align}

We consider all lipids, save cholesterol, to have the same area per lipid, 
$a=0.7$ nm$^2$, and take that of cholesterol to be $r_aa=0.4$nm$^2$ 
\cite{hung07,phillips71}. We assume that the areas of the two leaflets are 
equal which determines the ratio of molecules in the inner leaflet, 
$N^{(i)}$, to that in the outer leaflet, $N^{(o)}$
\begin{equation}
\frac{N^{(i)}}{N^{(o)}}=\frac{1-(1-r_a)y_{c_o}}{1-(1-r_a)y_{c_i}}.
\end{equation}
From this, we can express the mol fraction of the total amount of 
cholesterol in the bilayer in terms of $y_{C_i}$ and $y_{C_o}$:
\begin{equation}
x_C=\frac{y_{C_i}+y_{C_o}-2(1-r_a)y_{C_i}y_{C_o}}{2-(1-r_a)(y_{C_i}+y_{C_o})}.
\end{equation}

The free energy of the system consists of several parts. First there is 
the simple, regular-solution, free energy which can be written
\begin{align}
F_{rs}&=\int\ d^2r f_{rs}\nonumber \\ &=\int\ \frac{d^2r}{a}\left[ 
\frac{1}{(1-(1-r_a)y_{c_i})}f_1(\{y_i\})+\frac{1}{(1-(1-r_a)y_{c_o})}
f_2(\{y_i\})\right],\nonumber
\end{align}
where the free energies per molecule are
\begin{align}
f_{1}[T,y_{PS},y_{PE},y_{C_{i}}]
		 =& \, 6\epsilon_{PS-PE} y_{PS}y_{PE}\nonumber\\
		 &+6\epsilon_{PS-C} y_{PS}y_{C_{i}}
		 +6\epsilon_{PE-C} y_{PE}y_{C_{i}}\nonumber\\
&+k_BT(y_{PS}\log y_{PS} + y_{PE}\log y_{PE} + y_{C_{i}}\log y_{C_{i}}),
\nonumber
	 \text{ }\\
	f_{2}[T,y_{SM_{or}},y_{SM_{dis}},y_{PC},y_{C_{o}}]
		 = &\, 6\epsilon_{SM_{or}-PC} y_{SM_{or}}y_{PC}+ 6\epsilon_{SM_{dis}-PC} y_{SM_{dis}}y_{PC}\nonumber\\
		 &+6\epsilon_{SM_{or}-C} y_{SM_{or}}y_{C_{o}}+6\epsilon_{SM_{dis}-C} y_{SM_{dis}}y_{C_{o}}\nonumber\\
		 &+6\epsilon_{PC-C} y_{PC}y_{C_{o}}\nonumber
		 \\&+k_BT(y_{SM_{or}}\log y_{SM_{or}}  +y_{SM_{dis}}\log y_{SM_{dis}} \nonumber\\
		 & +
	 	 y_{PC}\log y_{PC} + y_{C_{o}}\log y_{C_{o}}).
\end{align}
To this we add the free energy associated with the deformation of the membrane 
which is described by deviations, $h$, from a flat configuration. There is 
the surface free energy
\begin{equation}
F_s=\int \frac{\sigma}{2}(\nabla h)^2\ d^2r,
\end{equation}
where $\sigma$ is the surface tension, and the bending energy
\begin{equation}
\label{bend}
F_b=\int\ \frac{\kappa}{2}(\nabla^2h-H_0)^2\ d^2r,
\end{equation}
where $\kappa$ is the bending modulus and $H_0$ is the spontaneous 
curvature. As in our previous paper \cite{giang14}, we take the latter to be due 
solely to the PE and to depend upon the cholesterol concentration in the 
inner leaflet
\begin{equation}
H_0=y_{PE}\left[H_{PE}-B\frac{y_{C_i}}{y_{min}}+\frac{B}{\lambda}
\left(\frac{y_{C_i}}{y_{min}}\right)^{\lambda}\right],
\end{equation}
with $B=0.05$, $y_{min}=0.3$ and $\lambda=8$. 
Note that the cross term in Eq. (\ref{bend}) couples the height fluctuations to
the composition-dependent spontaneous curvature. It is this term which can bring about modulated phases and/or microemulsions. 

The free energy includes the usual square-gradient term penalizing deviations from 
homogeneity,
\begin{equation}
F_p=\int\frac{1}{2}\sum_{i=1}^5b_i |\nabla y_i|^2\ d^2r.
\end{equation}
We take $b_i=k_BT.$
Lastly  we include a direct coupling between leaflets which promotes alignment 
of cholesterol-rich domains
\begin{equation}
F_c=\int f_c\ d^2r=-\int \Lambda(y_{C_i}-y_{PE})(y_{C_o}-y_{PC})\ d^2r.
\end{equation}
The coupling $\Lambda$ is not identical to the mismatch free energy defined previously \cite{may09,putzel11} and recently measured \cite{blosser15}, but we expect it to be similar within an order of magnitude. 

The total free energy is
\[F_{tot}=F_{rs}+F_s+F_b+F_p+F_c.\]

We utilize the two conditions
\begin{align}
&y_{SM_{ord}}+y_{SM_{dis}}+y_{PC}+y_{C_o}=1,\\
&y_{PE}+y_{PS}+y_{C_i}=1
\end{align}
to reduce from seven to five the number of independent mol fractions, and express the free energy in terms of them.

At equilibrium, a homogeneous phase is specified by these five  mol fractions, $y_i$.
They are determined by five conditions, as follows.
From the free energy, we calculate the chemical potentials of the two classes of SM.
Because the ordered and disordered SM can freely interchange, we  require that their chemical potentials be equal
\begin{equation}
\mu_{SM_{dis}}=\mu_{SM_{or}}.
\end{equation}
Similarly, as the cholesterol in the inner and outer leaflets can exchange freely, we require that their
chemical potentials, obtained from the free energy above, be equal,
\begin{equation}
\mu_{C_i}=\mu_{C_o}.
\end{equation}

The remaining three conditions are set by specifying the ratio of the 
total amount of SM to PC in the outer leaflet, the ratio of PS to PE in 
the inner leaflet, and the total mol fraction of cholesterol in the 
bilayer, $x_C$. These are known from experiment \cite{zachowski93}.

In order to determine the structure functions and correlation functions, 
we assume small deviations of the compositions from their equilibrium 
values, $y_i$, and small deviations $h$ of the membrane from a flat 
configuration. Extensions to a closed vesicle can be carried out 
\cite{kawakatsu93,taniguchi94,lavrentovitch16} but are unnecessary for our 
purposes. We expand the total free energy to second order in these 
deviations. In terms of the Fourier transforms $h(k)$ of $h({\bf r})$ and 
$y_i(k)$ of $y_i({\bf r})$ we obtain 
\begin{align} \label{fourier} 
F_{tot}&=\int\ 
d^2r\left[f_{rs}(\{y_i\})+f_c(\{y_i\})+\frac{\kappa}{2}H_o^2\right]\nonumber\\
         &+\frac{A^2}{(2\pi)^2}\int\ d^2k
         \left[\frac{1}{2}(\kappa k^4+\sigma k^2)h(k)h(-k)+
         \kappa k^2{\cal H}_0(k)h(-k)\right.\nonumber\\
         &\left.+\frac{1}{2}\sum_{i=1}^5b_ik^2y_i(k)y_i(-k)\right],
\end{align}        
where ${\cal H}_0$ is the Fourier transform of $H_0$ when expanded to 
first order in the compositions; i.e.
\begin{equation}
{\cal H}_0(k)=H_0(y_{PE},y_{c_i})\delta_{k,0}+\frac{\partial H_0}{\partial 
y_{PE}}\delta y_{PE}(k)+\frac{\partial H_0}{\partial y_{c_i}}\delta 
y_{c_i}(k).
\end{equation}
Upon minimizing $F_{tot}$ with respect to $h(k)$ we find
\begin{equation}
h(k)=-\frac{\kappa {\cal H}_0(k)}{\kappa k^2+\sigma},
\end{equation}
which we substitute into $F_{tot}$ to obtain, in Fourier space, an 
expression for it entirely in terms of the compositions $y_i(k)$. Lastly 
we expand this expression to second order in the $y_i(k)$ and obtain the 
second variation, $F_{tot}^{(2)},$ in matrix form
\begin{align}
&F_{tot}^{(2)}=\nonumber\\
&\frac{A^2}{(2\pi)^2}\int d^2k\{y_{PE}(k)y_{c_i}(k)y_{SM}(k)y_{c_o}(k)\} M
\{y_{PE}(-k)y_{c_i}(-k)y_{SM}(-k)y_{c_o}(-k)\}^T
\end{align}
where $y_{SM}=y_{SM_{dis}}+y_{SM_{or}}$ and $M$ is a $4\times 4$ matrix. 
From it, the matrix of structure factors follows: $S=M^{-1}.$ Correlation 
functions are obtained directly from the structure functions.

We now specify the properties of the membrane and of the interactions. We 
take the bending modulus to be $\kappa=44k_BT,$ appropriate to the plasma 
membrane \cite{evans83}, and the surface tension $\sigma=0.001k_B$T/nm$^2$ 
\cite{dai99}. The interactions are such that their ratios are, for the 
most part, taken from experiment \cite{almeida09}, and are the same as 
those we have used previously \cite{giang14}. 
We have also imposed a restrictive condition that the modulated lamellar
phase not be stable for any ratio of SM to PC  or PS to PE. Of course
the behavior of the plasma membtane for all such ratios is unknown, but
we impose this condition so as to avoid prejudicing a case for the
presence of a microemulsion by stabilizing a modulated phase at values
of lipid ratios near those of the actual plasma membrane.
This
requires that we take 
the absolute values of the interactions to be 
smaller than previously corresponding to temperatures somewhat 
greater than those of Table 1 of Ref.\cite{almeida09}. These 
interactions are, among components in the inner leaflet, 
$\epsilon_{PE-Ci}/k_BT=0.22$, $\epsilon_{PS,C_i}/k_BT=-0.05$, 
$\epsilon_{PE-PS}/k_BT=0,$ 
and in the outer leaflet $\epsilon_{PC-C_o}/k_BT=0.16$, 
$\epsilon_{SM_{ord}-C_o}/k_BT=-0.464,$ $\epsilon_{SM_{dis}-C_o}/k_BT=-0.23$, 
$\epsilon_{SM_{ord}-PC}/k_BT=0.24$, $\epsilon_{SM_{dis}-PC}/k_BT=0$. For the 
coupling between leaflets, we take $\Lambda=0.01k_BT$/nm$^2$ 
\cite{putzel11b}. The ratio of the compositions of SM to PC is taken from 
experiment, \cite{zachowski93} $(x_{SM_{ord}}+x_{SM_{dis}})/x_{PC}= 1.1,$ 
as is that of PS to PE $x_{PS}/x_{PE}=0.52.$ We take the total cholesterol 
composition of the bilayer to be $x_C=0.41$ \cite{vanmeer11}.

\section{Results}

\begin{figure}
\subfloat[The structure factor $S_{PE-PE}(k)$ peaks at $k=0.17$ 
$\text{nm}^{-1}$ with the corresponding wavelength of $\lambda = 37$ 
nm]{\includegraphics[width=0.45\textwidth]{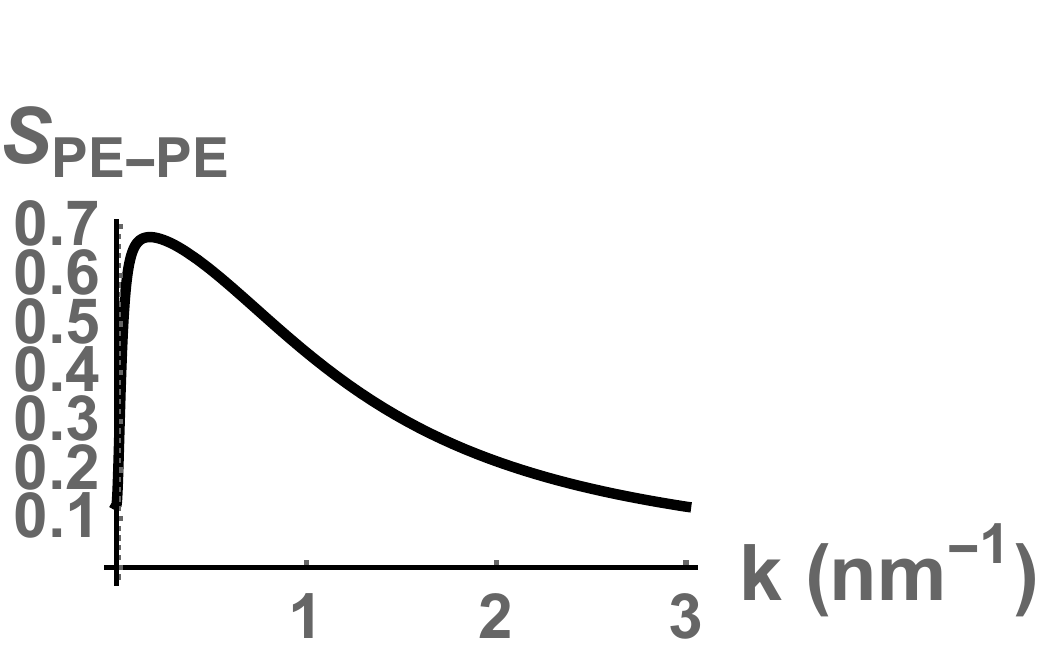}}
\hspace{0.5cm}
\subfloat[The correlation function $g_{PE-PE}(r)$ with correlation length 
of $\xi = 0.7$ nm]{\includegraphics[width=0.45\textwidth]{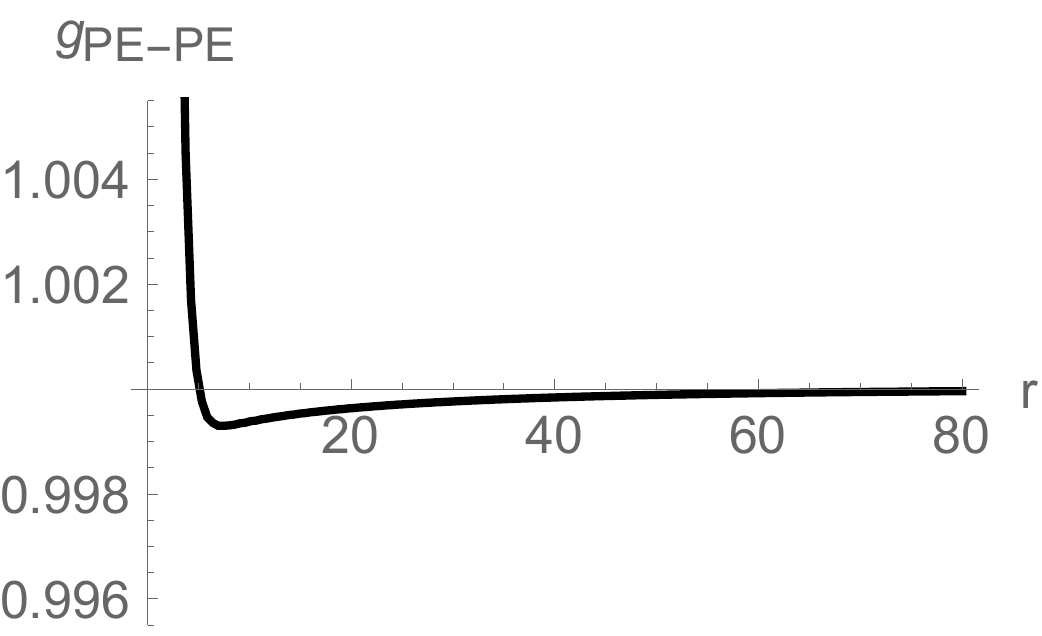}}
\caption[]{The correlation between fluctuations in the concentration of PE 
in the inner leaflet}
\label{fig:fig1}
\end{figure}

\begin{figure}
\subfloat[The structure factor 
$S_{SM-SM}(k)$]{\includegraphics[width=0.5\textwidth]{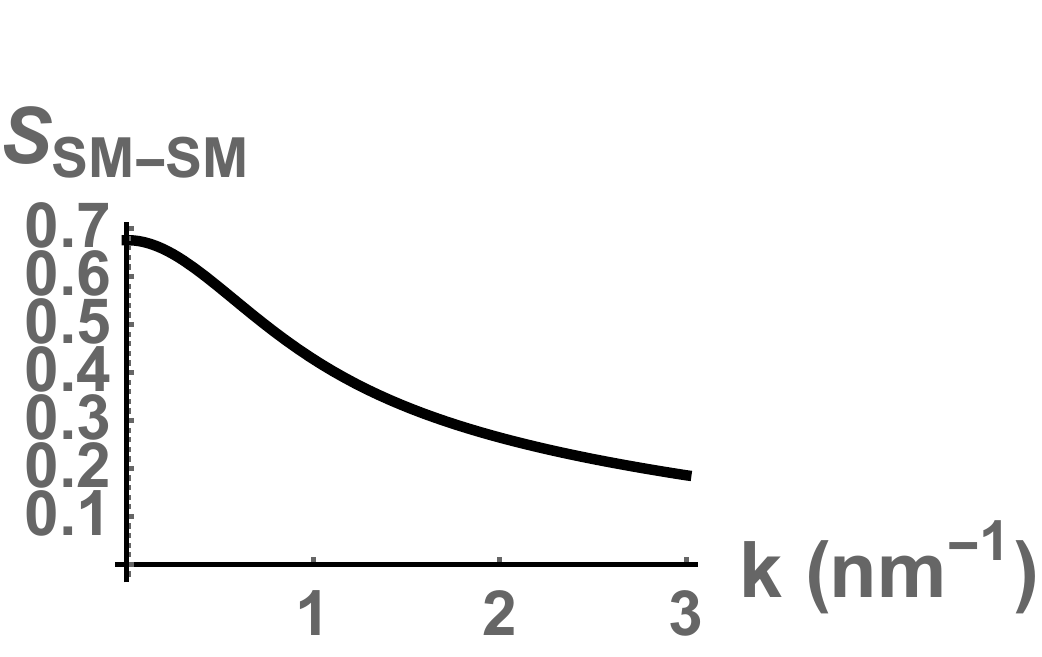}}
\subfloat[The correlation function 
$g_{SM-SM}(r)$]{\includegraphics[width=0.5\textwidth]{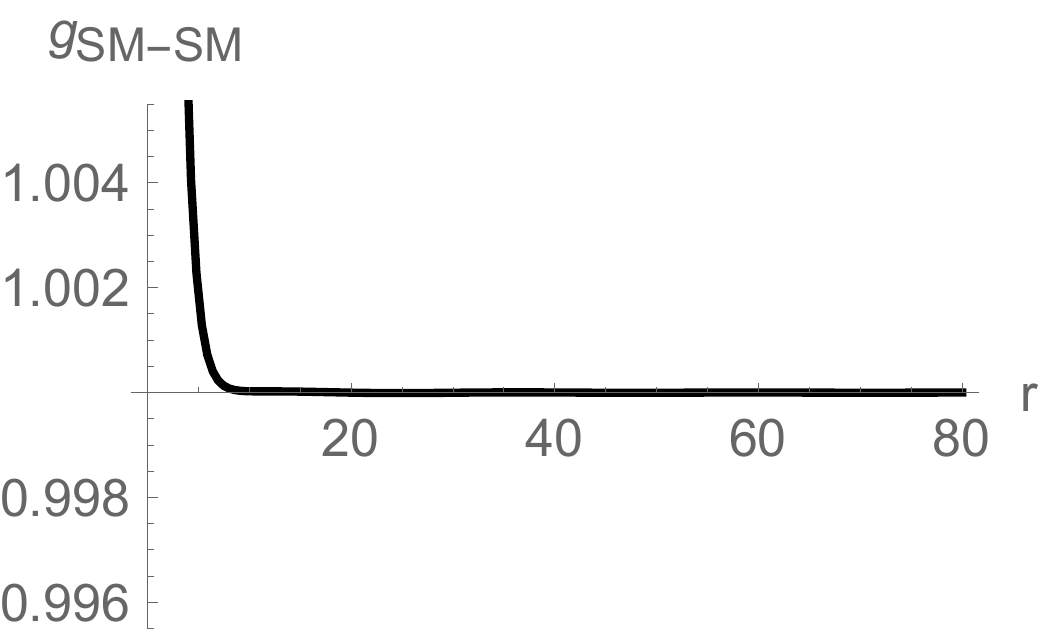}}
\caption[]{The correlation between fluctuations in the concentration of SM 
in the outer leaflet}
\label{fig:fig2}
\end{figure}

\begin{figure}
\subfloat[$S_{c_{i}-c_{i}}(k)$]{\includegraphics[width=0.5
\textwidth]{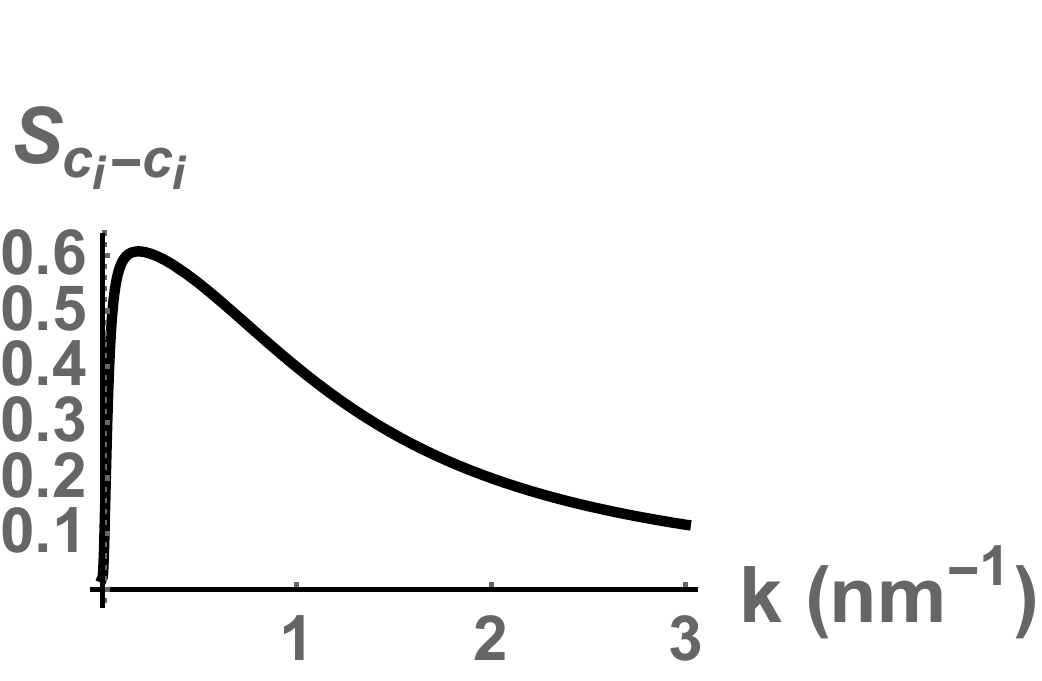}}
\subfloat[$S_{PE-c_{i}}(k)$]{\includegraphics[width=0.5\textwidth]{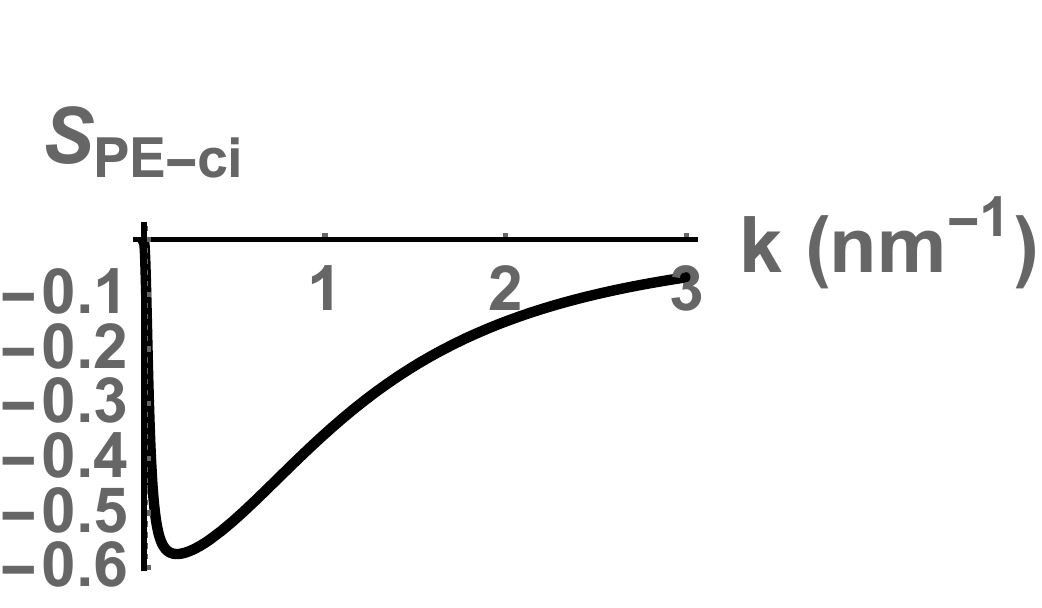}}
\caption[]{Two of the structure functions characterizing the inner leaflet.}
\label{fig:inner}
\end{figure}

\begin{figure}
\subfloat[$S_{c_{o}-c_{o}}(k)$]{\includegraphics[width=0.5\textwidth]{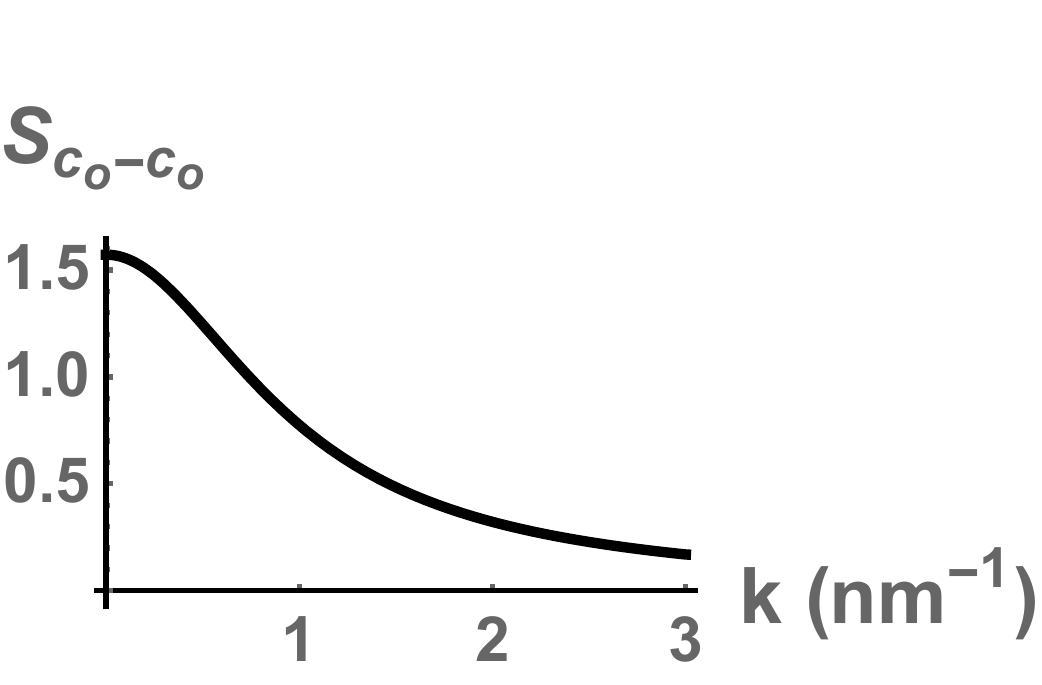}}
\subfloat[$S_{SM-c_{o}}(k)$]{\includegraphics[width=0.5\textwidth]{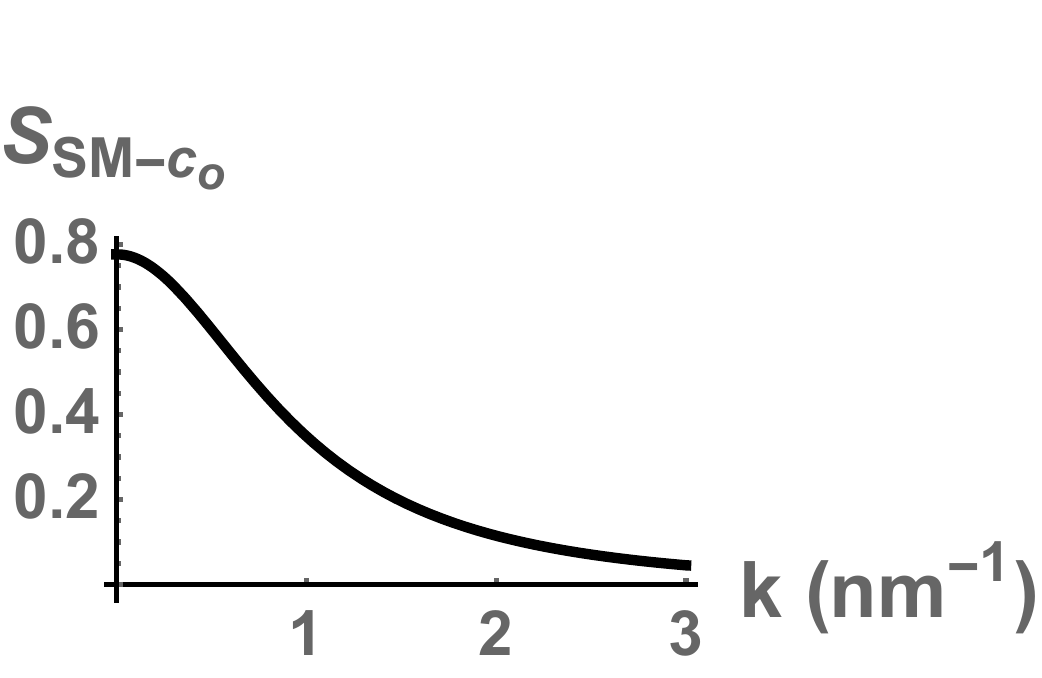}}
\caption[]{Two of the structure functions characterizing the outer leaflet}
\label{fig:outer}
\end{figure}

\begin{figure}
\subfloat[$S_{SM-c_{i}}(k)$]{\includegraphics[width=0.5\textwidth]{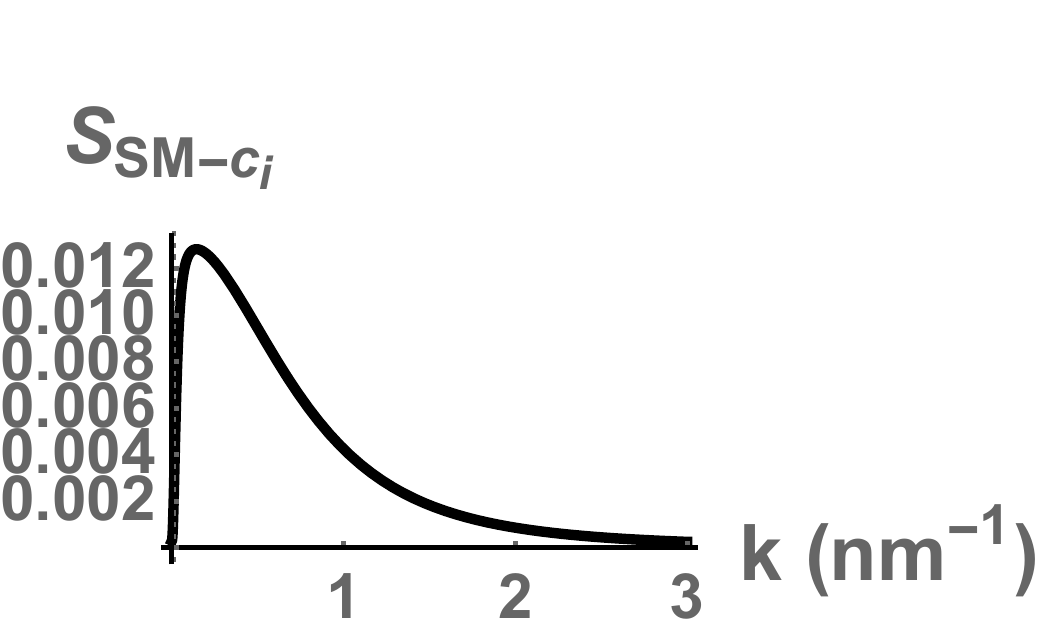}}
\subfloat[$S_{PE-c_{o}}(k)$]{\includegraphics[width=0.5\textwidth]{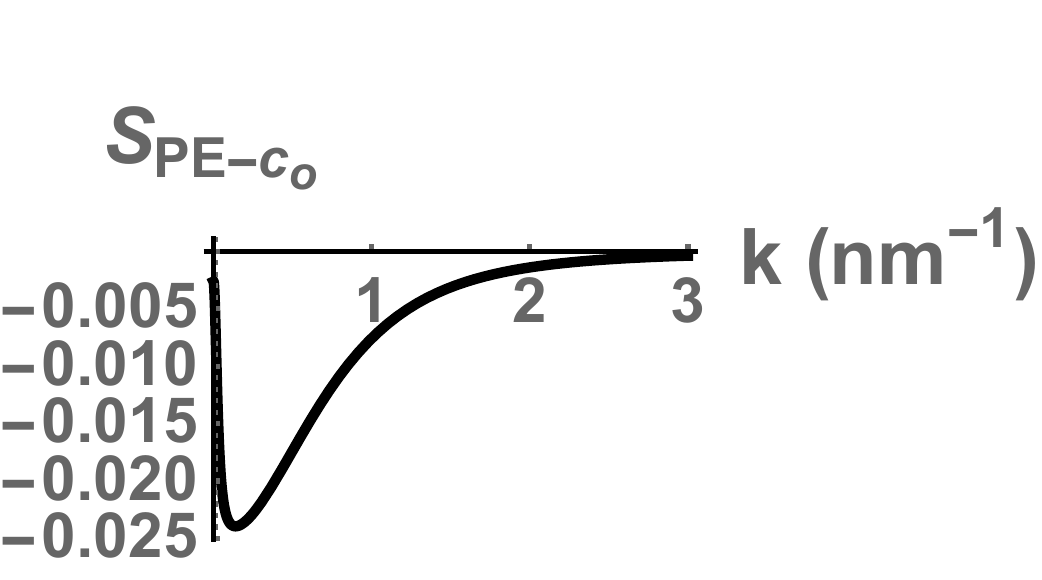}}
\caption[]{Two structure functions characterizing the correlation between 
compositions in the two leaflets.}
\label{fig:cross}
\end{figure}

We find that the components of the inner leaflet display characteristics 
typical of a microemulsion. In particular the structure factor 
$S_{PE-PE}(k),$ shown in Fig. \ref{fig:fig1}(a), which displays the 
correlation between fluctuations in the concentration of PE in the inner 
leaflet, is characterized by a peak at a non-zero wave vector $k=0.17$ 
nm$^{-1}$ which corresponds to a wavelength of $\lambda=37$ nm.
The Fourier transform of this function, the PE-PE correlation function 
$g(r)$, is shown in Fig. \ref{fig:fig1}(b). It displays no oscillations 
because the correlation length is $\xi=0.7$ nm, much shorter than the 
wavelength $\lambda$. Nonetheless, it is notable that the correlation 
function does not decay as a simple exponential, but approaches its 
asymptotic value from below, a remnant of the exponentially damped 
oscillatory behavior. In contrast to the inner leaflet, the outer one does 
not display the characteristics of a microemulsion, but rather that of a 
normal fluid. For example, the structure function $S_{SM-SM}(k)$, shown in 
Fig. \ref{fig:fig2}(a), displays a peak at zero wave vector, and the 
correlation function, shown in Fig. \ref{fig:fig2}(b), decays to zero from 
above with the same correlation length $\xi=0.7$ nm. Such a coupling of a 
microemulsion in one leaflet and a normal fluid in the other has been 
discussed previously \cite{hirose09,hirose12}. If the coupling $\Lambda$ 
between leaflets is increased by an order of magnitude, we find that the 
microemulsion in the inner leaflet is conveyed to the outer one as 
evidenced by the fact that $S_{SM-SM}(k)$ now displays a peak at non-zero 
wave vector.
The structure functions arising from fluctuations in any pair of 
components can be obtained. We show in Fig. \ref{fig:inner} the 
$S_{c_i-c_i}$ and $S_{PE-c_i}$ functions. The former shows microemulsion 
behavior of cholesterol in the inner leaflet, and anticorrelated behavior 
of PE and cholesterol in the inner leaflet. Similarly Fig. \ref{fig:outer} 
shows $S_{c_o-c_o}$ and $S_{SM-c_o}$. These show that the outer leaflet is 
a normal fluid and that the SM and cholesterol in the outer leaflet are 
correlated as expected. Finally in Fig. \ref{fig:cross} we display 
$S_{SM-c_i}$ and $S_{PE-c_o}$ which shows the fluctuations in SM and 
cholesterol in the inner leaflet to be correlated while those of PE and 
cholesterol in the outer leaflet to be anticorrelated. In Table 
\ref{table:lipc} we show the concentrations of the various components in 
the two leaflets. The percentage of the total cholesterol which is in the 
inner leaflet is 61\% in accord with most experiments which show that the 
cholesterol is either evenly divided between leaflets 
\cite{muller02,lange82}, or more abundant in the inner 
leaflet\cite{brasaemle88,schroeder91,wood90,igbavboa96,mondal09}.

\begin{table}[h]
\centering
\begin{tabular}{| lc  | lc |}
	\hline
	$y_{SM_{o}}$	& 0.18	&$x_{SM_{o}}$		&0.08\\
	$y_{SM_{d}}$	&0.18	&$x_{SM_{d}}$		&0.08\\
	$y_{PC}$		&0.32	&$x_{PC}$		&0.16\\
	$y_{c_{o}}$	&0.32	&$x_{c_{o}}$		&0.16\\
	$y_{PS}$		&0.18	&$x_{PS}$		&0.09\\
	$y_{PE}$		&0.35	&$x_{PE}$		&0.18\\
	$y_{c_{i}}$	&0.47	&$x_{c_{i}}$		&0.25\\
	\hline
\end{tabular}
\caption[]{Lipid composition of the two leaflets is shown on the left as 
mol fraction in the leaflet, and on the right as mol fraction in the 
bilayer.}
\label{table:lipc}
\end{table}

\begin{figure}
\subfloat[$S_{PE-PE}(k)$ peaks at $k=0.19$ $\text{ nm}^{-1}$ 
]{\includegraphics[width=0.5\textwidth]{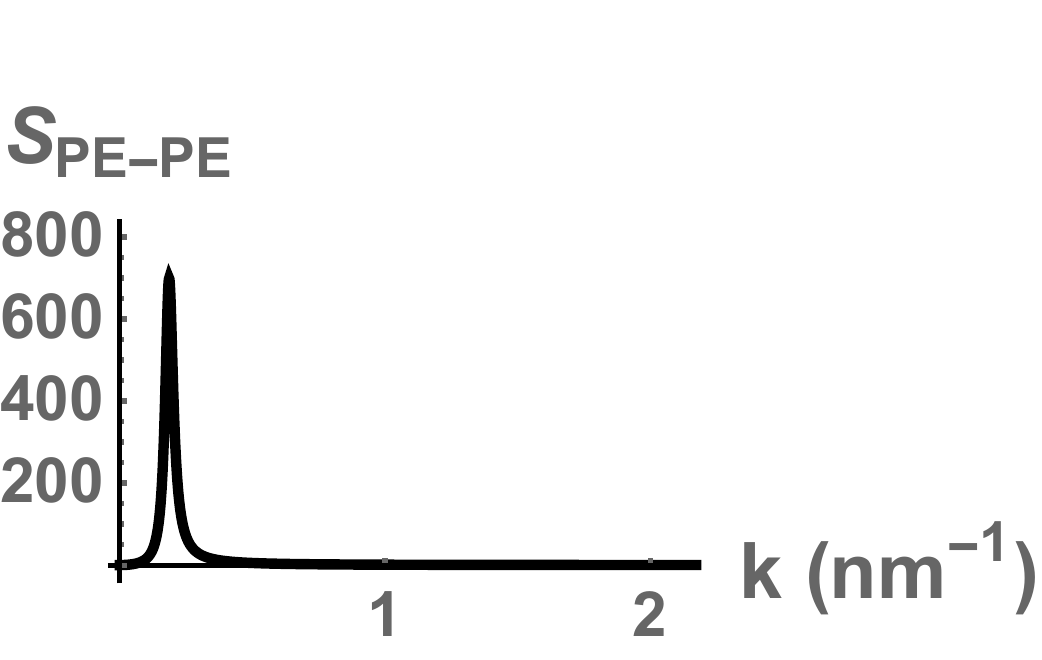}}
\subfloat[$g_{PE-PE}(r)$, $\xi_{PE-PE} = 55$ 
nm]{\includegraphics[width=0.5\textwidth]{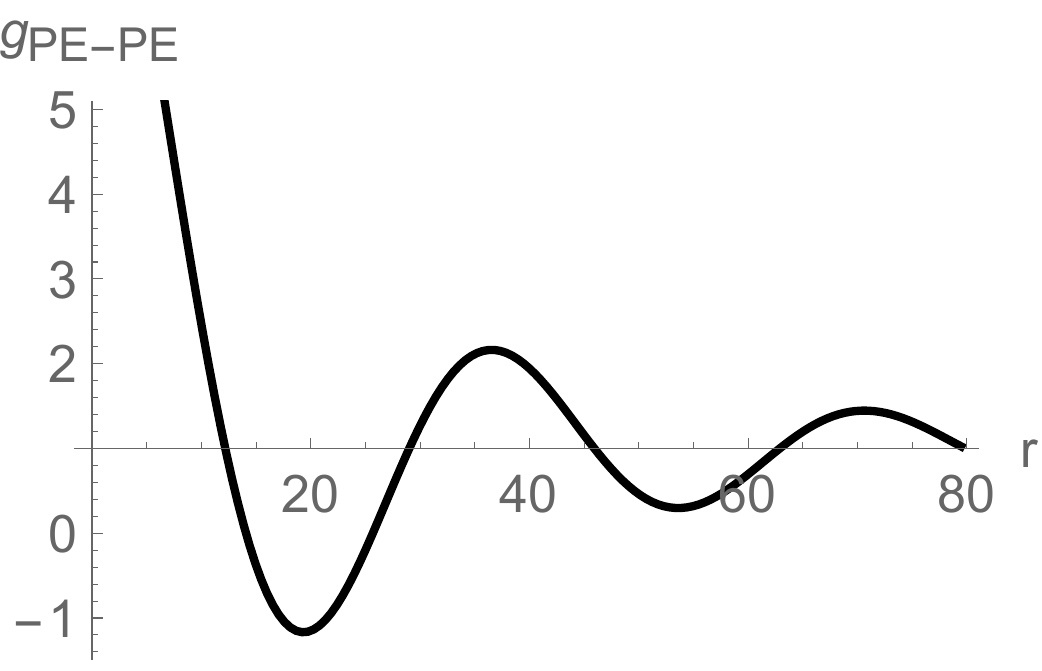}}
\subfloat[$S_{SM-SM}(k)$ peaks at $k = 0.19 \text{ 
nm}^{-1}$]{\includegraphics[width=0.5\textwidth]{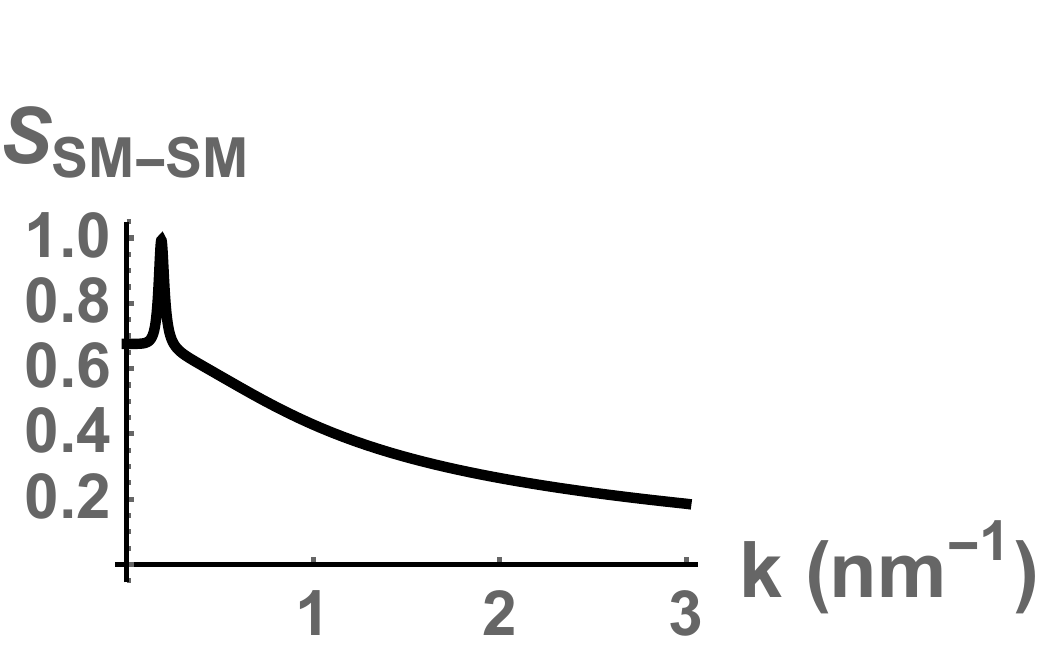}}
\subfloat[$g_{SM-SM}(r)$, $\xi_{SM-SM} = 41.5$ 
nm]{\includegraphics[width=0.5\textwidth]{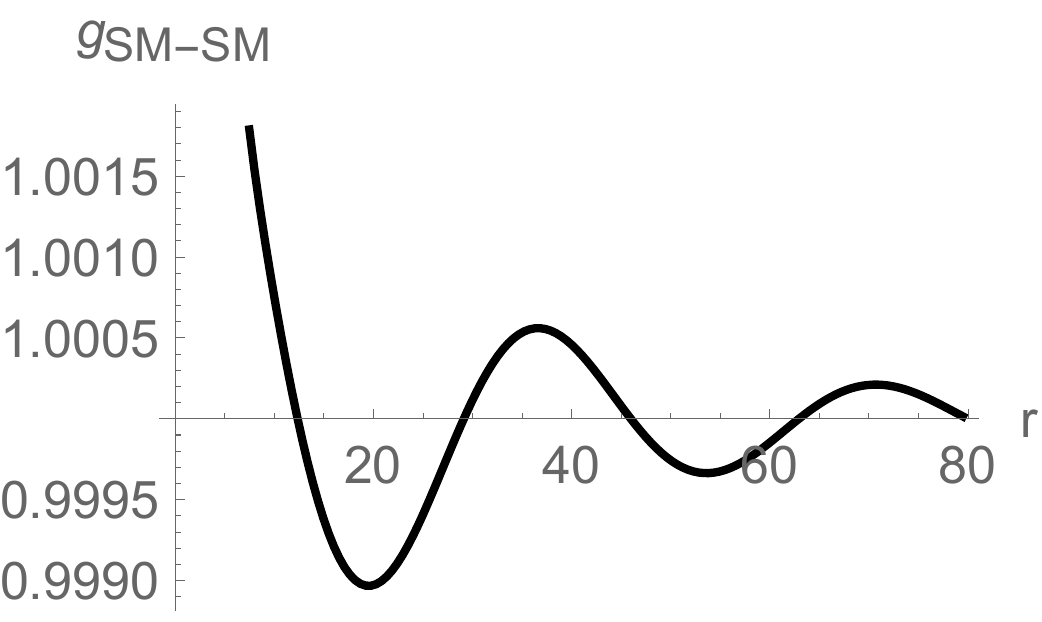}}
\caption[]{Exponentially damped oscillations in the correlation functions 
when $\beta = 1.0165$}
\label{fig:osci}
\end{figure}

That the inner leaflet is a microemulsion is due to the coupling of the 
membrane curvature to the membrane spontaneous curvature, dominated by 
that of PE. Thus the appearance of the microemulsion in the inner leaflet 
depends upon the strength of this coupling. To consider this further, we 
have multiplied by a parameter $\beta$ the term $\kappa k^2{\cal 
H}_0(k)h(-k)$ in Eq. (\ref{fourier})  that couples the height and 
composition fluctuations. We find that the appearance of the microemulsion 
in the inner leaflet is quite robust. The strength of the coupling as 
measured by $\beta$ would have to be reduced by more than two orders of 
magnitude for the inner leaflet to behave like an ordinary fluid as 
indicated by a peak at zero wave number in $S_{PE-PE}(k).$ Thus there is a 
very large range of values of the coupling strength over which the inner 
leaflet displays the properties of a microemulsion. On the other hand, we 
note that with a value of spontaneous curvature appropriate to PE the 
microemulsion is near its limit of stability.  With an increase of $\beta$ 
from its value of unity by about 1.7\%, the correlation functions in both 
leaflets, shown in Fig. \ref{fig:osci}, clearly display exponentially 
damped oscillatory behavior because the correlation lengths have increased 
to 41.5 and 55 nm, larger than the wavelength of about 33 nm. An increase 
in the coupling strength by 2\% brings about a transition to a modulated 
phase.

\section{Conclusions}
We have employed a model of an asymmetric plasma membrane with SM, PC, and 
cholesterol in the outer leaflet and PE, PS, and cholesterol in the inner 
leaflet. The membrane curvature is coupled to the concentration of PE in 
the inner leaflet via a cholesterol-dependent spontaneous curvature, one 
which reproduces the distribution of cholesterol between leaflets observed 
in several experiments. We find that the model predicts that the inner 
leaflet is almost certainly a microemulsion, but most probably a weak one 
with the correlation length, of order of a few nanometers or less, being 
much smaller than the wavelength of the spatial variations, of order 37 
nm. Consequently little of the latter are seen. Whether this microemulsion 
propagates to the outer leaflet depends upon the strength of the coupling 
between leaflets.

The results of this model for the origin of ``rafts'' differ significantly 
from those of others.  In particular the model states that such 
inhomogeneities, identified as arising from a microemulsion, originate in 
the inner leaflet. Furthermore because the microemulsion is weak, the 
model predicts that such inhomogeneities would not be present in a 
membrane consisting solely of lipids. However the response of the inner 
leaflet to external perturbations, due to non-lipid components, proteins 
perhaps, is largest at a distance of the order of 37 nm from the site of 
the perturbation. The origin of this length is clear in the model, and can 
be traced to the physical properties of the membrane itself via its 
bending rigidity and surface tension. The propagation of this response to 
the outer leaflet is enhanced by coupling between the leaflets.

The strength of the microemulsion, that is the degree to which spatial 
variations would be evident, depends in large part on the local 
spontaneous curvature of the membrane, which points to the importance of 
membrane composition.  Indeed we believe that the mechanism explored here 
provides a plausible explanation for the modulated phases observed in the 
yeast plasma membrane \cite{toulmay13} and other systems, both {\em in 
vitro}. \cite{konyakhina11,goh13,stanich13} and {\em in vivo} 
\cite{lavrentovitch16}.  Given the close connection between modulated 
phases and microemulsions, it follows that these systems are very likely 
to display the latter as well.

\section{Author Contributions}
HG designed and performed research, and analyzed data. MS designed
research, and wrote the paper.

\section{Acknowledgments} 
We are grateful to Sarah Keller and Lutz Maibaum and their groups for stimulating 
interactions.  This work was supported in part by the National Science 
Foundation under grant No. DMR-1203282.

\end{document}